\documentclass[twoside,final]{pro12}

\usepackage[ansinew]{inputenc}
\usepackage{amsmath}
\usepackage{graphicx}
\usepackage{array}
\usepackage{SIunits}
\usepackage{lscape}
\usepackage{multirow}
\usepackage{float}
\usepackage{natbib}
\usepackage{hyperref}
\usepackage{xcolor}

\newenvironment{correction}{}

\head{Mauduit et al.}{Exoplanetary radio emission prediction}	

\begin{document}

\title{PALANTIR: an updated prediction tool for exoplanetary radio emissions}	
\author{E. Mauduit\adress{\textsl LESIA, Observatoire de Paris, Universit\'{e} PSL, Sorbonne Universit\'{e}, Universit\'{e} Paris Cit\'{e}, CNRS, 92190 Meudon, France }$\,\,$, 
J.-M.  Grie{\ss}meier\adress{\textsl Laboratoire de Physique et Chimie de l'Environnement et de l'Espace (LPC2E) Universit\'{e} d'Orl\'{e}ans/CNRS, Orl\'{e}ans, France} $^{,}$\adress{Observatoire Radioastronomique de Nan\c{c}ay (ORN), Observatoire de Paris, Universit\'{e} PSL, Univ
Orl\'{e}ans, CNRS, 18330 Nan\c{c}ay, France}$\,\,$,
P. Zarka$^{1,3}$,
J.D. Turner\adress{\textsl Department of Astronomy and Carl Sagan Institute, Cornell University, Ithaca, NY, USA} $^{,}$\adress{\textsl NHFP Sagan Fellow}
}


\maketitle

\begin{abstract}
    In the past two decades, it has been convincingly argued that magnetospheric radio emissions, of cyclotron maser origin, can occur for exoplanetary systems, similarly as solar planets, with the same periodicity as the planetary orbit. These emissions are primarily expected at low frequencies \citep[usually below 100 MHz, c.f.][]{Farrell1999JGR,Zarka2007PSS}. 
    The radio detection of exoplanets will considerably expand the field of comparative magnetospheric physics and star-planet plasma interactions \citep{Hess2011AA}. 
    We have developed a prediction code for exoplanetary radio emissions,  
    PALANTIR: 
    ``Prediction Algorithm for star-pLANeT Interactions in Radio''.
    %
    This code has been developed for the construction of an up-to-date and evolutive target catalog, based on observed exoplanet physical parameters, radio emission theory, and magnetospheric physics embedded in scaling laws. It is based on, and extends, previous work by \citet{Griessmeier2007AA}.
    Using PALANTIR, we prepared an updated list of targets of interest for radio emissions. Additionally, we compare our results with previous studies conducted with similar models \citep{Griessmeier2017PRE8}. 
    For the next steps, we aim at improving this code by adding new models and updating those already used. 
\end{abstract}

\section{Introduction}

The purpose of PALANTIR (``Prediction Algorithm for star-pLANeT Interactions in Radio'') is to provide the exoplanetary radio community with a way to estimate fundamental parameters such as the radio flux at Earth $\Phi_{mag}$ and the maximum frequency, $f_c^{max}$.
These predictions are useful to both plan and optimize observations, and also to compare observations to predictions.
This new code is an updated version of previous work by \cite{Griessmeier2007AA}, with some improvements regarding the models used, especially for the estimation of planetary magnetic moment. 
In addition to 
the models already taken into account in \cite{Griessmeier2007AA}, PALANTIR also includes the magnetic moment estimate based on \cite{ReinersChristensen2010AA}. 
Two different lists of targets were obtained using these two sets of models. 

The code has been built to be user friendly. As input, it uses two mandatory input files. A third imput file is optional:
\begin{itemize}
    \item[-] the input catalog: a database with all the required parameters for targets (for example obtained from \url{exoplanet.eu}), 
    \item[-] a configuration file, in which the user can select the models to be used and set some constants,
    \item[-] (optional) a catalog with updated values: a complementary file in addition to main database. This database is maintained manually (based on recent literature), and includes parameters which override those of the input catalog.
    For the moment, this file is not yet used.
\end{itemize}

As output, PALANTIR indicates, for each of the targets of the input catalog, the estimated parameters for potential radio-emissions, such as :

\begin{itemize}
    \item[-] the magnetic moment of the planet, $\mathcal{M}$,
    \item[-] the planetary field at the surface above the pole, $B_p$, 
    \item[-] the stellar wind parameters at the planet, $v_{\text{eff}}$, $n_e$ and $T_{corona}$, 
    \item[-] the inter-planetary magnetic field at the planet, $B_{imf}$, 
    \item[-] the maximum cyclotron frequency at the planet, $f_c^{max}$,
    \item[-] and the radio flux at the observer, $\Phi_{mag}$.
\end{itemize}
The user can then set up the constraints specific to his study and obtain a sorted list of targets. For practical purposes, we estimate that all targets emit 
between 0 Hz and $f_c^{max}$;
for this reason, we only keep results with $f_c^{max}$ above a critical value and then sort by flux density.
As minimum value for $f_c^{max}$, we currently use 5 MHz. This value was chosen based on the 
ionospheric cut-off (10 MHz), but also takes into account typical uncertainties of our estimates
(approximately a factor 2-3 in $f_c^{max}$, see \citet{Griessmeier2007PSS}).

\section{Method}

\subsection{Radio-magnetic law}

As mentioned, the code is based on previous work. It mostly relies on the radio-magnetic law  \citep{Zarka2007PSS,Saur2013AA,Zarka2018AA}, suggesting that the radio power of the emission is proportional to the input magnetic power at the planet:

\begin{equation}
    P_{radio} = \eta_{radio} \epsilon P_{input,mag}
    \label{eq1}
\end{equation}

where $\epsilon$ is the fraction of the input power that is dissipated within the magnetosphere and $\eta_{radio}$ is the fraction of this power that is converted into radio-emission \citep{Zarka2007PSS}.
\correction{Equation (\ref{eq1}) ignores a correction factor which depends on the Mach number \citep[e.g. Eq. 8 of][]{Zarka2007PSS}. For low Mach numbers, this will slightly overestimate the radio power, but  still gives the correct order of magnitude.}

The product of both proportionality constants in Eq.~(\ref{eq1}) can be evaluated by comparison to Jupiter. \correction{As a matter of fact, it has been shown that the star-exoplanet system is, in some ways, analogue to Sun/Jupiter and Jupiter-Io/Ganymede systems \citep{Zarka2018AA}. Indeed, radio emissions at Jupiter can be powered by either the interaction between the solar wind and its magnetosphere, or by the interaction between Jupiter's magnetosphere and its satellites. In their work, \citet{Zarka2018AA} showed that Auroral, Io-induced and Ganymede-induced non-thermal radio emissions at Jupiter follow the same radio-magnetic scaling law as the gas giants emissions powered by solar wind, in the solar system. Thus, it seems reasonable to use Jupiter as a reference for our radio flux estimations.} In this work, we take the average power during periods of high \correction{planetary} activity as a reference value: $P_{radio,J} = 2.1 \times 10^{11}$ W \citep{Zarka2004JGR}.\correction{This value corresponds to the contribution of all radio emissions, from $0.4$ MHz to $30$ MHz, powered by a flow-obstacle interaction at Jupiter and where a high activity period is defined by having radio emissions more than $50$\% of the time in the time interval considered.} See \citet{Zarka2004JGR} for further details. 

%
It has been suggested \citep{Zarka2001ASS} and then confirmed by \citet{Zarka2018AA} that the magnetic contribution to the total input power is the most important one.
In this work, the input magnetic power is given by:

\begin{equation}
    P_{input}^{mag} \propto v_{\text{eff}} B_{\perp}^2 R_S^2
    \label{eq2}
\end{equation}

where $v_{\text{eff}}$ is the effective velocity of the stellar wind (i.e., in the frame of the planet), $B_{\perp}$ denotes the component, perpendicular to the stellar wind flow, of the interplanetary magnetic field, and $R_S$ is the standoff distance of the planetary magnetosphere. Therefore, we need to estimate both stellar wind and magnetospheric parameters. 

\subsection{Stellar wind model}

Similarly to the work of \citet{Griessmeier2007AA}, we use the Parker solar wind model to estimate the parameters of the stellar wind. \correction{The Parker model gives the following equation on the velocity: }
\begin{equation}
    \left( \frac{v(d)}{v_{crit}}\right)^2 - 2\ln\left( \frac{v(d)}{v_{crit}}\right) = 4\ln\left(\frac{d}{r_{crit}}\right) + 4\frac{r_{crit}}{d} - 3
    \label{eq:parker}
\end{equation}

\correction{where $d$ is the distance from the star, $r_{crit} = \displaystyle \frac{m_pGM_*}{4 k_B T_{corona}}$ the critical radius (beyond which the flow is supersonic), $v(d)$ is the velocity of the stellar wind at $d$ and $v_{crit} =v(r=r_{crit})=
 \displaystyle \sqrt{\frac{k_B T_{corona}}{m_p}}$ is the critical velocity.}
\correction{In order to solve this equation and find the velocity of the stellar wind at the planet, we proceed as follows. First, we compute the velocity of the solar wind at $1$ AU, at the age of the star $t_*$, using the equation given in \citet{Griessmeier2007AA}:
\begin{equation}
    v_{sun}(1\text{AU},t) = v_0 \left( 1+\frac{t}{\tau}\right)^{-0.43}
\end{equation}
where $v_0 = 3971$ km.s$^{-1}$ and $\tau = 2.56\times 10^7$ yrs.
Then, we estimate the temperature of the corona such that $v(d=1\text{AU}) = v_{sun}(1\text{AU},t_*)$. When we have $T_{corona}$, we can then solve Eq.~(\ref{eq:parker}) and obtain the stellar wind velocity at the planet and deduce from that the parameters we need, such as the effective velocity $v_{eff} = \sqrt{v^2 + v_{orb}^2}$. Here, $v_{orb}$ is the planetary orbital velocity.}
\correction{For further details, see \citet{Griessmeier2007AA}.}

\correction{This model present some limitations, mainly on the age and rotation period of the star. Indeed, a high rotation rate will lead to a higher stellar wind velocity than the one given by the Parker model. However, as shown in \citet{Preusse2005AA}, for a star with a rotation period similar to the Sun, there is no significant difference between the Parker model and the more complex model of \citet{Weber1967ApJ}.
For more rapid rotation, a difference appears and can reach a factor 2 in stellar wind velocity for rotation period of about $3$ days. With this, our stellar wind model is limited to main sequence stars of age $\ge 0.7$ Gyr, which corresponds to rotation periods $\ge 7$ days, see \citet{Griessmeier2007PSS} for further details.}

\correction{The stellar wind magnetic field is determined as follows: first, each component of the interplanetary magnetic field ($B_{imf,r}$, $B_{imf,\phi}$) is computed by propagating the magnetic field at the stellar surface to $1$ AU. Then, similarly as \citep{Griessmeier2007AA} we can deduce the perpendicular stellar magnetic field (IMF) $B_{\perp}$: }
\begin{equation}
    B_{\perp} = \sqrt{B_{imf,r}^2 + B_{imf,\phi}^2} \times \lvert \sin{(\alpha - \beta)} \rvert
\end{equation}
\correction{where, $\alpha = \arctan \left(B_{imf,\phi}/B_{imf,r}\right)$ and $\beta = \arctan (v_{orbit}/v)$.}
\correction{Moreover, the stellar magnetic field at the surface, $B_\star$ is computed using Eq.~(23) of \citet{Griessmeier2007AA} :}
\begin{equation}
    \frac{B_*}{B_{sun}} = \frac{P_{sun}}{P_*}
\end{equation}

\correction{where $P_{sun} = 25.5$ days, $B_{sun} = 1.435 \times 10^{-4}$ T and, $P_* \propto \left(1+t_*/\tau \right)^{0.7}$. 
The limitations of these models are mostly due to the estimation of the magnetic field values at the surface. In the future, this could be improved by using these values directly where available, rather than relying on the stellar rotation.}

\subsection{Magnetic moment models}
\label{section2.2}
The main update of this code in comparison with \citet{Griessmeier2007AA} resides in the models used to estimate the magnetic moment of the planets. Previously, the magnetic moment was computed using four different models, all based on the effect of the rotation rate of the planet on the dynamo region. They are given by: 
\begin{itemize}
    \item Mizutani with slow convection \citep{Mizutani1992ASR}: $\mathcal{M} = \sqrt{\omega\rho_c  / \sigma} \,\ r_c^{3}$,
    \item Mizutani with moderate convection \citep{Mizutani1992ASR}: $\mathcal{M} = \rho_c^{1/2}\omega^{3/4} \sigma^{-1/4} r_c^{7/2}$,
    \item Busse \citep{Busse1976PEPI}: $\mathcal{M} = \rho_c^{1/2} \omega \,\ r_c^{4}$,
    \item Sano \citep{Sano1993JGG}: $\mathcal{M} = \rho_c^{1/2} \omega \,\ r_c^{7/2}$,
\end{itemize} 

where $\rho_c$ and $r_c$ are, respectively, the density and the radius of the dynamo region, $\omega$ is the rotation rate of the planet and $\sigma$ is the electrical conductivity of the planet (here we assume it is the same as for Jupiter, see \citet{Griessmeier2007AA}). 
\correction{Similarly to \citet{Griessmeier2007AA}, the parameters of the dynamo region ($\rho_c$ and $r_c$) are obtained by solving the Lane-Emden equation. This consist in finding the core radius for which the density is high enough so that the transition to the liquid-metallic state can occur. In this work we used $\rho_{transition} = 700$ kg.m$^{-3}$ \citep[see][for further details]{Griessmeier2007AA}.}
In this work we still use these four models but, the mean value is computed differently. Previously, it was obtained by taking the geometrical mean value between the minimum and the maximum value obtained with the different models : $\mathcal{M}=\sqrt{\mathcal{M}_{max} \times \mathcal{M}_{min}}$. Now, we take the geometrical mean value of the magnetic moments obtained with all models (here $n=4$). Thus, the magnetic moment is given by :

\begin{equation}
    \mathcal{M} = \left(\prod_{i=1}^n \mathcal{M}_i \right)^{1/n}
    \label{eq:moment_MSB}
\end{equation}

In addition, we implemented a new model for the magnetic moment based on the work of \citet{ReinersChristensen2010AA}, which we analyze independently. In this model, the magnetic moment mostly depends on the mass of the planet, its age (assumed to be similar to the host star) and its luminosity, which drives the internal convection. 
 The mean magnetic field strength at the dynamo is given by:

\begin{equation}
    B_{dyn} = 0.48 \times \left( \frac{M_p L_p^2}{R_p^7}\right)^{1/6} \,\ [G]
    \label{eq:bdyn}
\end{equation}

where $M_p$ is the mass of the planet, $L_p$ its luminosity and $R_p$ its radius; these values are normalized to the Sun. 
\correction{The luminosity of the planet is determined by interpolating over planetary mass $ M_p$, and then over the stellar age $t_*$, based on tables in \citet{Baraffe2008AA}.}
We need to distinguish two cases: giant planets and brown dwarfs. For brown dwarfs, the top of the dynamo region is assumed to be at the surface, therefore $B_{dip} = B_{dyn}$. 
However, for giant planets, i.e. $M_p \leq 13 M_J$ ($M_J$ denotes Jupiter's mass),
the top of the dynamo region is within the planet, thus the relevant parameter is the dipole magnetic field strength at the equator. It is given by:

\begin{equation}
    B_{dip}^{eq} = \frac{B_{dyn}}{2\sqrt{2}} \left(1 - \frac{0.17}{M_p/M_J}\right)^3
    \label{eq6}
\end{equation}

Then, 
the magnetic moment is then derived by using: $ \mathcal{M} = B_{dip}^{pol} R_p^3$ [$\mathcal{M}_J$], where $B_{dip}^{pol} = 2B_{dip}^{eq}$. 
For this study, we focused on targets with $M_p \leq 13 M_J$. See \citet{ReinersChristensen2010AA} for further details on this model.

\correction{For the moment, PALANTIR does take irradiation of the planet (especially for Hot Jupiters) into account for the planetary radius $R_p$ \citep[following the procedure of][]{Griessmeier2007AA}. This irradiation, however, is not taken into account for the planetary luminosity $L_p$. The difference, however, is comparatively small: for an orbital distance of 0.45 AU and planetary masses $\ge0.1$ $M_J$,  \citet{Baraffe2008AA} find a difference in planetary luminosity of less than a factor of 3. This translates to a difference of a factor 1.4 for the magnetic moment and maximum emission frequency, i.e., less than our typical error \citep[approximately a factor 2-3 in $f_c^{max}$, see][]{Griessmeier2007PSS}.}

\subsection{Radio-emissions parameters}

The parameters which we are most interested in are the maximum frequency of the probable emission and the flux density of the emission at Earth. The maximum frequency of the emission is given by \citep{Farrell1999JGR}:

\begin{equation}
    f_c^{max} = \frac{e B_p}{2\pi m_e} \,\ \,\ \text{[MHz]}
\end{equation}

where $e$ is the elementary charge, $m_e$ is the electron mass and $B_p$ defines the polar magnetic field of the planet at its surface.
computed from its magnetic moment. This frequency corresponds to the local electron cyclotron frequency.

The second parameter of interest is the radio flux of the emission seen by an observer on Earth. It is given, for example, by \citep{Griessmeier2007PSS}:

\begin{equation}
    \Phi_{mag} = \frac{P_{radio}}{\Omega s^2 \Delta f},
\end{equation}

where $s$ is the distance between the observer and the emitter, $\Omega$ defines the solid angle of the beam of the emitted radiation, here we take $\Omega = 1.6$ sr \citep{Zarka2004JGR} and $\Delta f$ is the bandwidth of the emission, here we assume it to be equivalent to the maximum cyclotron frequency, $f_c^{max}$ \citep{Griessmeier2007PSS}.

\section{Results}

\begin{figure}[h!]
\centering
\includegraphics[width=1\textwidth]{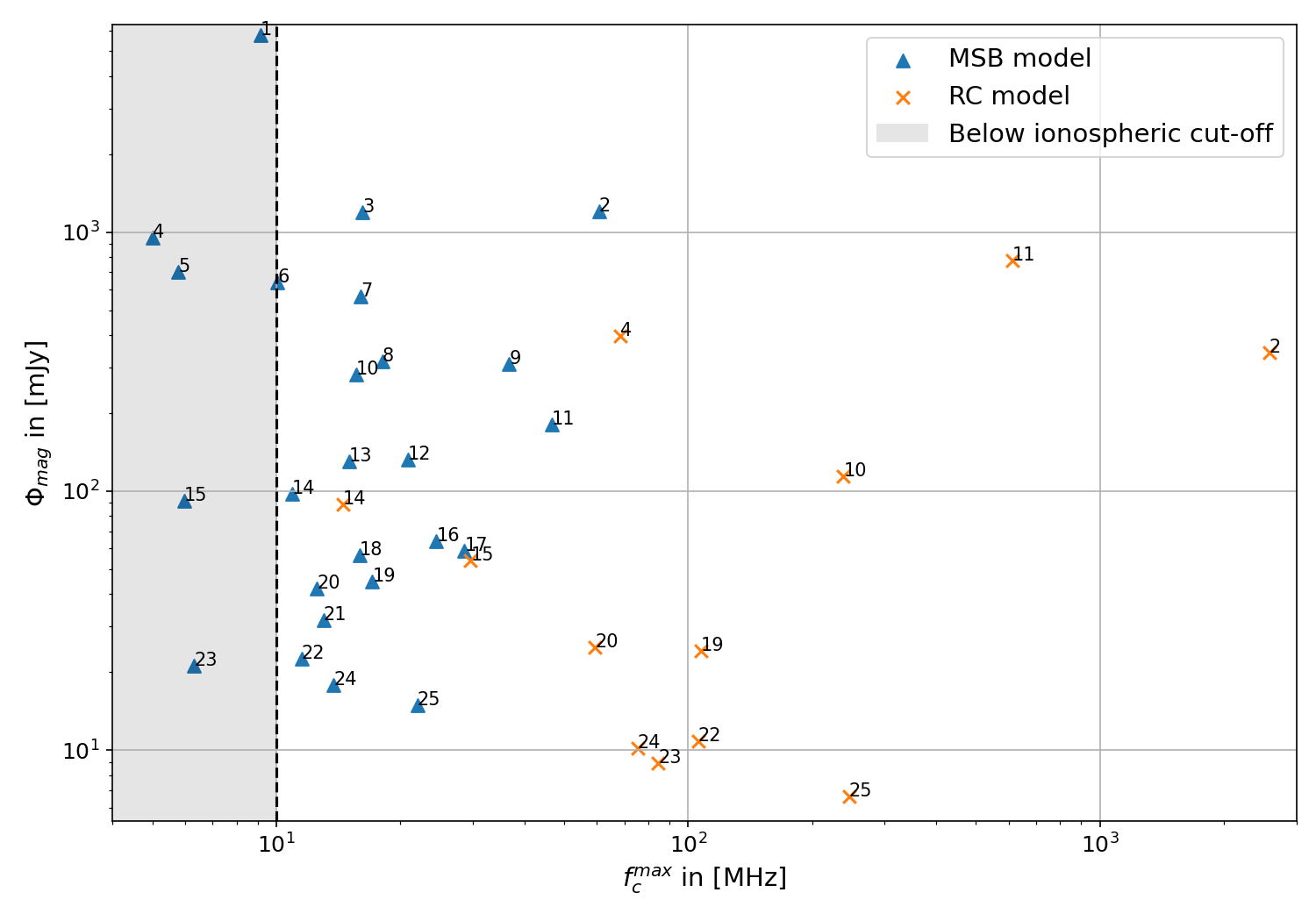}
\caption{\correction{Maximum emission frequency and expected radio flux for 25 extrasolar planets. The grey zone below $f_c^{max} = 10$ MHz represents the frequencies that are below the ionospheric cut-off, thus not observable from the ground. The values are given for two different sets of models for the magnetic moment: $MSB$ refers to the combination of the magnetic field models of Mizutani, Busse and Sano. $RC$ refers to the model of \citet{ReinersChristensen2010AA}.}}
\label{fig:figure1}
\end{figure}

We apply the code PALANTIR to the data base \url{exoplanet.eu}  (retrieved on $23^{rd}$ of February $2023$).
We only kept planets with \correction{$f_c \geq 5$ MHz }, which is a little below the ionospheric cut-off, to take into account potential uncertainties on the computed parameters. We then extracted the 25 targets with the highest radio flux densities. 
 In Table~\ref{table1} we give the parameters obtained for these 25 targets for the two sets of magnetic moment models used. We also give other parameters that may be of interest, such as the planetary mass and radius $M_p$ and $R_p$, periastron distance $a$, stellar age $t_*$, and the distance to the observer $s$. Those parameters are the ones that contributes the most in the final estimation of frequency and radio flux. \correction{In Figure~\ref{fig:figure1}, we show the expected radio flux as a function of the maximum emission frequency, for both magnetic moment models.}

\begin{table}[h!]
\footnotesize
\centering
\begin{tabular}{|c|c||c|c|c|c|c||c|c||c|c|}
        \hline
		N$^\circ$ & Target & $M_p$ & $R_p$ & $a$ & $t_*$ & $s$ & $f_{max}^{MSB}$ & $\Phi_{mag}^{MSB}$ & $f_{max}^{RC}$ & $\Phi_{mag}^{RC}$ \\
         & & [$M_J$] & [$R_J$] & [AU] & [Gyr] & [pc] & [MHz] & [mJy] & [MHz] & [mJy]
  \\\hline
		1 & GJ 1214 b & 0.03 & 0.24 & 0.01 & 6.00 & 13.00 & 9.14 & 5752 & - & -\\
        2 & HD 41004 B b & 21.25 & 1.06 & 0.02 & 1.56 & 43.03 & 60.7 & 1199 & \correction{2581} & \correction{343.28}\\
        3 & GJ 367 b & 0.002 & 0.06 & 0.01 & 5.20 & 9.41 & \correction{16.2} & \correction{1191} & - & -\\
        4 & HD 189733 b & 1.14 & 1.14 & 0.03 & 0.60 & 19.30 & 5.00 & 951.96 & \correction{68.44} & \correction{397.91}\\
        5 & TRAPPIST-1 b & 0.003 & 0.10 & 0.01 & 7.60 & 12.10 & 5.77 & 700.67 & - & -\\
        6 & GJ 486 b & 0.009 & 0.12 & 0.02 & 5.20 & 8.07 & 10.05 & 637.6 & - & - \\
        7 & GJ 1252 b & 0.01 & 0.11 & 0.01 & 5.20 & 20.38 & \correction{16} & \correction{562.8} & - & -\\
        8 & LTT 3780 b & 0.01 & 0.12 & 0.01 & 5.20 & 21.98 & 18.10 & 315.89 & - & -\\
        9 & Kepler-42 c & 0.01 & 0.07 & 0.01 & 4.50 & 38.70 & 36.72 & 309.43 & - & -\\
        10 & tau Boo A b & 5.84 & 1.06 & 0.04 & 2.52 & 15.60 & 15.63 & 282.25 & \correction{238.48} & \correction{113.79}\\
        11 & WASP-18 b & 10.43 & 1.17 & 0.02 & 0.63 & 100.00 & 46.60 & 181.29 & \correction{612.06} & \correction{76.84}\\
        12 & 55 Cnc e & 0.03 & 0.17 & 0.02 & 10.20 & 12.34 & 20.87 & 132.38 & - & -\\
        13 & TOI-1634 b & 0.02 & 0.16 & 0.01 & 5.20 & 35.27 & 15.01 & 130.59 & - & -\\
        14 & PSR J2322-2650 b & 0.92 & 1.31 & 0.01 & 5.20 & 230 & 10.90 & 97.84 & \correction{14.48} & \correction{89.01} \\
        15 & HD 143105 b & 1.39 & 1.2 & 0.04 & 5.2 & 48.7 & 5.96 & 91.68 & \correction{29.59} & \correction{53.74} \\
        16 & TOI-431 b & 0.01 & 0.11 & 0.01 & 5.20 & 32.61 & 24.44 & 63.95 & - & -\\
        17 & TOI-1238 b & 0.01 & 0.11 & 0.01 & 0.80 & 70.64 & \correction{28.63} & \correction{58.6} & - & -\\
        18 & L 168-9 b & 0.01 & 0.12 & 0.02 & 5.20 & 25.15 & 15.94 & 56.71 & - & -\\
        19 & HD 80606 b & 3.94 & 0.92 & 0.03 & 7.63 & 58.40 & \correction{17.1} & \correction{44.6} & \correction{107.44} & \correction{24.13}\\
        20 & Qatar-2 b & 2.49 & 1.25 & 0.02 & 5.00 & 182.32 & 12.54 & 41.93 & \correction{59.42} & \correction{24.96}\\
        21 & GJ 9827 b & 0.02 & 0.14 & 0.02 & 5.20 & 30.3 & 13.0 & 31.67 & - & - \\ 
        22 & HATS-24 b & 2.44 & 1.49 & 0.02 & 0.88 & 510.00 & 11.50 & 22.60 & \correction{105.7} & \correction{10.79}\\
        23 & WASP-140 b & 2.44 & 1.44 & 0.03 & 1.60 & 180 & 6.31 & 21.2 & \correction{84.6} & \correction{8.92} \\
        24 & KELT-16 b & 2.75 & 1.42 & 0.02 & 2.9 & 365 & 13.76 & 17.88 & \correction{75.31} & \correction{10.15} \\
        25 & HAT-P-20 b & 7.25 & 0.87 & 0.04 & 6.7 & 70 & 22.02 & 14.84 & \correction{245.9} & \correction{6.64} \\
          \hline
\end{tabular}
\caption{Predicted radio emission frequencies and flux densities at Earth for the 25 best targets. The values are given for two different sets of models for the magnetic moment: $MSB$ refers to the combination of the magnetic field models of Mizutani, Busse and Sano. $RC$ refers to the model of \citet{ReinersChristensen2010AA}. In this table, $M_p$ is the planetary mass and $R_p$ is the planetary radius both with respect to Jupiter. We also give the periastron distance $a$, the stellar age $t_*$, the distance from Earth \correction{$s$}. Finally, we give the maximum frequency of the predicted emission, $f_{max}$, and the corresponding radio flux at Earth, $\Phi_{mag}$, for both models. }
\label{table1}
\end{table}

Both sets of models we use have a minimum mass, below which the planet does not have a significant magnetic moment, and radio emission vanishes.
As it can be seen in Table~\ref{table1}, for the Reiners-Christensen model, some values are missing. This is explained by the definition of the magnetic field given by Eq.~(\ref{eq6}), where the magnetic field become negative, thus non-physical, when the planetary mass is $M_p \le 0.17 M_J$. Below this mass, the planet is assumed to be too small to have a dynamo region, therefore no magnetic field can be created. Similarly, for the MSB model (combining the scaling laws of Mizutani, Busse and Sano using eq.~(\ref{eq:moment_MSB})), the limitation on the mass also comes from the ability of the planet to have a dynamo region. If the planet is too small, it does not have a metallic core, and no magnetic field is created.

Moreover, the \correction{Reiners-Christensen} model has already been studied by \citet{Lynch2018MNRAS}, but with some differences. We use different database, different values for some constants (for Jupiter's radius for example) and slightly different ways to compute some parameters.

\section{Discussion}

In a previous PRE proceeding, a similar list of target was provided \citep{Griessmeier2017PRE8}. The differences between our predictions for the same targets (in Table~\ref{table1}) can be explained by the fact that since $2017$, the database has been updated, with more precision on some crucial parameters such as the planetary mass, $M_p$, the planetary radius $R_p$, the orbital period, $\omega_{orb}$. Moreover, as explained in Section~\ref{section2.2}, the magnetic moment $\mathcal{M}_p$ of the planet is computed differently even
when using the same models as this previous work. The differences are within the error margin we expect for our estimations. As presented in \citet{Griessmeier2007PSS}, the uncertainty on the radio flux $\Phi_{mag}$ strongly depends on the uncertainty of the stellar age $t_*$, which is about ~$50\%$. The uncertainty on the maximum frequency of emission $ f_c^{max}$ relies on the uncertainty on the magnetic moment of the planet $\mathcal{M}_p$, which is estimated to be about a factor of two. The values and uncertainties are different for the \correction{Reiners-Christensen} model.

Even with differences in the codes, some targets, such as $\tau$~Bo\"{o}tis, always come up as targets of choice in almost every work, including this one, on radio emission predictions \correction{\citep{Griessmeier2007AA,ReinersChristensen2010AA,Griessmeier2017PRE8,Lynch2018MNRAS,Ashtari2022ApJ}}.
For this reason, we think that 
these studies 
can be a helpful tool for observers in order to select the targets they want to observe. Indeed, $\tau$~Bo\"{o}tis for example is the subject of several past and ongoing 
observational campaigns. It is also the source of a tentative signal \citep[see the work of][]{Turner2021AA}, which is currently being followed-up \citep[see the work of][]{Turner23PRE9}.

Finally, one can see in Table ~\ref{table1} that except for a few targets, most of the predicted frequency are in the low radio frequency range, $f_c^{max} \leq 20$ MHz. Therefore, for observational studies telescopes with a high sensitivity at low frequencies are needed. This study has been performed with the aim to create an evolving list of target for ground-based observational studies. The new radiotelescope NenuFAR \citep{Zarka2018URSI}, a pathfinder of SKA, will provide exceptionally high sensitivity at the lowest radio frequencies observable from the ground, between the ionospheric cutoff (10 MHz) and the FM radio band (87 MHz). One of the long-term project of NenuFAR is dedicated to the search for exoplanetary radio-emissions, and 55 targets have already been observed. The data analysis is ongoing.

\section{Conclusions and Perspectives}

With PALANTIR, we present an updated, more flexible and more user-friendly implementation of the radio prediction code that was first used by \citet{Griessmeier2007AA}. We aim at making it available for the community, with a ready-to-use interface and a proper documentation. The user will only have to provide a database, to specify the models to use and the criteria for target selection. 
For example, we are in the process of applying PALANTIR to the complete exoplanet census, based on \url{www.exoplanet.eu}; details and results will be shown in a further more complete paper.".

The modularity and flexibility of PALANTIR will allow us to easily add new models, for example concerning the magnetic field model, the interaction powering the radio emission,  or to take into account the quenching of planetary radio emission by a dense, ionized atmosphere as expected for low-mass hot Jupiter-like planets \citep[see, for example][]{Griessmeier23PRE9}.

\section{Acknowledgements}
This work has made use of the Extrasolar Planet Encyclopaedia (exoplanet.eu) maintained by J. Schneider \citep{Schneider2011AA}. PZ acknowledges funding from the ERC under the European Union's Horizon 2020 research and innovation programme (grant agreement no. 101020459 - Exoradio).
\correction{We thank the anonymous referees for helpful and constructive suggestions.}

\newcommand{\newblock}{}
\bibliographystyle{mnras}
\bibliography{bibliography.bib}

\end{document}